\documentclass{ws-ijgmmp}
\usepackage{amssymb}
\usepackage{diagbox}
\usepackage{amsmath}
\usepackage{amsfonts}
\usepackage{bm}
\usepackage{cancel}
\usepackage{comment}

\newcommand\be{\begin{equation}}
\newcommand\ee{\end{equation}}

\newcommand{\bea}{\begin{eqnarray}}
\newcommand{\eea}{\end{eqnarray}}

\allowdisplaybreaks[4]

\begin{document}

\markboth{Artyom V. Astashenok, Alexander S. Tepliakov}
{Tsallis holographic dark energy model with event horizon cutoff in modified gravity}

%
\catchline{}{}{}{}{}
%

\title{Tsallis holographic dark energy model with event horizon cutoff in modified gravity
}

\author{ARTYOM V. ASTASHENOK}

\address{Immanuel Kant Baltic Federal University, Nevskogo, 14\\
Kaliningrad 236041, Russia\\
aastashenok@kantiana.ru }

\author{ALEXANDER S. TEPLIAKOV}

\address{Immanuel Kant Baltic Federal University, Nevskogo, 14\\
Kaliningrad 236041, Russia\\
atepliakov@kantiana.ru }

\maketitle

\begin{history}
\received{(Day Month Year)}
\revised{(Day Month Year)}
\end{history}

\begin{abstract}

We considered the Tsallis holographic dark energy model in frames of Nojiri-Odintsov gravity with $f(R)=R+\lambda R^2-\sigma{\mu}/{R}$. For IR cutoff event horizon is taken. The cosmological evolution of such universe is investigated for various initial conditions and values of parameters. The dependence of the Hubble parameter $H$ from time in the future has an oscillations. It is shown that for $\mu \neq 0$ appearance of singularities are typical and the time up to these singularities can be relatively small from cosmological viewpoint. The singularity is associated with the zero of second deribative of $f(R)$ on $R$. It is interesting to note that these models can describe observational data from Ia supernovae astrophysics and dependence of the Hubble parameter from redshift $z$ at least not worse than canonical $\Lambda$CDM model. 

\end{abstract}

\keywords{Dark energy; holographic energy; modified gravity.}

\section{Introduction}

The classical theory of gravitation, which describes the gravitational interaction, was developed by Isaac Newton in the second half of the 17th century. Based on Kepler's empirical laws, Newton showed that the observed motion of the planets is under the action of a central force and that this central gravitational force leads to motion along second-order curves (circle, ellipse, parabola, hyperbola), with the central body being in focus. Despite the fact that the theory well described the observed motions of planets and objects in the solar system, it suffered from a number of problems, such as unexplained rapidity (the gravitational force was transferred infinitely fast and through empty space), gravitational paradox (in an infinite universe with Euclidean geometry and non-zero average matter density, the gravitational potential takes infinite value everywhere). By the end of the 19th century, the French astronomer Urbain Le Verrier, who developed the theory of the motion of Mercury, discovered a discrepancy between the theory and observations: the perihelion of Mercury's orbit shifted somewhat faster than it followed from the theory. These problems prompted the search for new theories describing the gravitational interaction.

The solution was proposed by Albert Einstein, who developed in 1915 general theory of relativity (GTR), which solved the problems of classical gravitation theory and postulated gravitation as a manifestation of spacetime geometry. Considering evolution of Universe Einstein assumed that it is stationary. For this he included the $\Lambda$ constant in the equations. But Edwin Hubble showed that the universe expands. After this Einstein eventually admitted that $\Lambda$ was a major mistake of his life. 

In 1998 an analysis of the luminosity of Type Ia supernovae in distant galaxies showed that the Universe expands with an acceleration \cite{1}, \cite{2}. To explain this acceleration we need a new substance in the Universe - so called dark energy, distributed in space with a high degree of homogeneity, having low density and interacting with ordinary matter only through gravity. The simplest explanation for dark energy is nothing else tnan Einstein Cosmological Constant (or vacuum energy) $\Lambda$. Such a model of the Universe is called the $\Lambda$CDM model \cite{LCDM-1,LCDM-2,LCDM-3,LCDM-4,LCDM-5,LCDM-6,LCDM-7}. This model satisfies to observational data with high precision and is currently the standard cosmological model. However, there are more complex descriptions for dark energy phenomena, consistent with observations. In particular, many models of holographic dark energy \cite{Miao, Qing-Guo, Qing-Guo-2} have been proposed. These models are based on the holographic principle \cite{Wang,3,4,5} derived from black hole thermodynamics. According to this principle all physical quantities inside the Universe, including the dark energy density, can be described by setting of some quantities on its boundary. Tsallis' generalization of the Boltzmann-Gibbs entropy expression for black holes \cite{Tsallis, Tsallis-2} led to a new class of dark energy model, namely the Tsallis holographic dark energy model (THDE) \cite{Tavayef,Moosavi,AA}. 

There are many papers devoted to THDEs. For example, various infrared (IR) cutoffs, including particle horizon, Ricci horizon and Grande-Oliveros (GO) cutoffs have been studied \cite{Zadeh}. Authors of \cite{Ghaffari,Jawad} studied the cosmological consequences of THDE in the framework of Brans-Dicke gravity and modified Brans-Dicke gravity. Most general Tsallis HDE was introduced in \cite{Nojiri:2019skr}. We considered THDE in cosmology on the Randall-Sandrum brane \cite{AA}, and recently investigated the THDE model with inclusion of a possible interaction between matter and dark energy \cite{AA2}.

Another possible way to describe cosmological acceleration is modified gravity \cite{reviews1, reviews2, reviews3, reviews4, reviews5, book, reviews6}. Modified gravity models assume that the universe is accelerating due to the deviations of real gravity from GTR on cosmological scales. $F(R)$ gravity is known as the simplest modification of GTR \cite{Harko, reviews3}. In this theory, the Einstein-Hilbert action is changed by replacing of the Ricci scalar $R$ on some function $f(R)$.

It is interesting to consider Tsallis holographic dark energy on background of $f(R)$ gravity by the following reasons. $f(R)$ gravity can explain some features of inflation i.e. early cosmological acceleration. And maybe late cosmological acceleration caused by another source namely holographic energy. Therefore consideration of holographic dark energy on non-GTR background has a sense. Also some new effects from holographic dark energy in a case of modified gravity may appear.  

In \cite{Ens}, various $f(R)$ gravity models have been considered with the inclusion of THDE with simple IR cutoff $L~1/H$ where $H$ is the Hubble parameter. We investigate the more realistic case (in GTR) of THDE with event horizon as IR cutoff in  model of gravity with $f(R)=R+\lambda R^2 - \sigma {\mu}/{R}$. In next section basic cosmological equations for universe contained THDE in $f(R)$ gravity are presented. In GTR cosmological equations contain second derivatives of scale factor $d(t)$ but in $f(R)$ gravity third derivative appears. This allows to construct solutions with similar behaviour in past but which split in future. Then we considered in detail solutions with various parameters of THDE and $\lambda$, $\mu$. For considered model of gravity the main feature of cosmological evolution is that Hubble parameter oscillates with time near the dependence corresponding to pure GTR. Also singularity in future appears. This singularity corresponds to zero of second derivative of $f''(R)$. The time before singularity can be relatively small. We performed brief analysis of observational constrains and demonstrated that such models in principle are reliable from astrophysics viewpoint.   

\section{Basic equations}

In $f(R)$ gravity action is written in the following form (hereafter we use the natural system of units, in which $8\pi G=c^2=1$): 
\bea
S = \frac{1}{2}\int f(R)\sqrt{-g}\,d^4x + S_M,
\eea
where $g$ is the determinant of the spacetime metric $g_{\mu \nu}$, and $S_m$ is the matter action, $f(R)$ is a continuous, twice differentiable function of its argument. Varying the action by the metric, we obtain the equations for gravitational field:
\bea
R_{\mu\nu}f'(R)-\frac{1}{2}g_{\mu\nu}f(R)-\nabla_\mu\nabla_\nu f'(R)+g_{\mu\nu}\Box f'(R)= \kappa\, T_{\mu\nu}. 
\eea
Here prime means the derivative with respect to scalar curvature $R$, $\nabla_\mu$ is the covariant derivative with respect to the coordinate $x_\mu$, $\Box\equiv\nabla_\mu\nabla^\mu $ and $T_{\mu\nu} $ is the energy-momentum tensor defined by the relation
\be
T_{\mu\nu}=-\frac{2}{\sqrt{-g}}\frac{\delta S_M}{\delta g^{\mu\nu}}. 
\ee
Next, consider the spatially flat universe described by the Friedman-Lemetre-Robertson-Walker (FLRW) metric:
\bea
ds^2 = -dt^2 + a^2(t)\left[dr^2 + r^2d\theta^2 + r^2sin^2\theta d\phi^2\right],
\eea
where $a(t)$ is the scale factor. For this metric the cosmological equations can be written as follows: 
\be
H^2 = \frac{1}{3f'(R)}\left(\rho + \frac{Rf'(R)-f(R)}{2} - 3H\dot{R}f''(R)\right),\label{friedmann00}
\ee
\be
\dot{\rho }+3H(\rho +p)=0. \label{friedmann01}
\ee
Here $H = {\dot{a}}/{a}$ is the Hubble parameter, $\rho = \rho_{de}+\rho_{m}$ is the total energy density in the Universe (we neglect the contribution of radiation, which is essential only at early times of the cosmological evolution), where $\rho_{de}$ is the dark energy density and $\rho_{m}$ is the matter density. The continuity equation (\ref{friedmann01}) is valid for each of the components separately in the absence of interaction between them. 

The scalar curvature for the FLRW metric is expressed as follows:
\begin{equation}
\label{eq:R}
R = 6\left ( \dot{H} + 2H^2\right ).
\end{equation}

The density of the Tsallis holographic dark energy is given by the expression: 
\begin{equation}
\label{eq:rho_de}
\rho_{de} = \frac{3C^2}{L^{\alpha}}, 
\end{equation}
in which $\alpha = 4-2\gamma$ and $C = \mbox{const}$. For $\gamma=1$ we obtain the simplest holographic model. For IR cutoff $L$ one can choose the Hubble horizon, event horizon, the particle horizon or its combination. We investigate in details the case when $L$ is the event horizon:
\begin{equation}
\label{eq:8}
L = a\int_{t}^{t_{s}}\frac{dt}{a}.
\end{equation}
Here $t_{s}$ is the time of possible future singularity. For $\gamma=1$ this choice to be the most relevant in terms of consistency with the observational data. Then the total energy density is 
\begin{equation}
\label{eq:10}
\rho =\rho_{de} + \rho_m = \frac{C^2}{\tilde{L}^\alpha a^\alpha} + \frac{1-\Omega}{a^3},
\end{equation}
where 
$$\tilde{L}= \dfrac{L}{a} = \int_{t}^{t_{s}}\dfrac{dt}{a}$$ 
and 
$$\rho_m = \dfrac{1-\Omega}{a^3}$$ 
is the matter density. Parameter $\Omega$ has a simple sence of dark energy fraction in total energy budget of universe at current moment of time $t=0$ (without loss of generality $a=1$ for $t=0$ is assumed). We take $\Omega=0.72$  similar to the standard cosmological model. {Note that Eq. (\ref{eq:rho_de}) can be considered as some limit of general holographic HDE was firstly introduced in \cite{Nojiri:2005pu}. In this work the following class of model was proposed}:
$$
\rho_{de} = \frac{3C^2}{L^{2}_{\Lambda}},
$$
$$
{L_\Lambda}=\frac{ \left( \frac{t_s B\left( 1+h_0,1-h_0 \right)}{L_p + L_f}\right)^{1/h_0}}{ h_0
\left\{ 1 + \left(\frac{t_s B\left(1+h_0,1-h_0\right)}{L_p + L_f}\right)^{1/h_0}\right\}^2}
\ ,\quad h_0>0\ .
$$
{Here $B$ is beta-function and $L_{p}$ and $L_{f}$ are particle and future horizon correspondingly. If $t_{s}$ is large (for cosmological model without singularity $t_{s}=\infty$) we can neglect first term in denominator and obtain simply that}
$$
{L_\Lambda} \sim (L_{p}+L_{f})^{1/h_{0}}.
$$
{Therefore model (\ref{eq:rho_de}) is a particular case of general model from \cite{Nojiri:2005pu} with replacement $L_{p}+L_{f} \rightarrow L_{f}$ and $t_{s}\rightarrow \infty$.}  

We consider the gravity model obtained by combining the models proposed by Starobinsky and Carroll-Duvvuri-Trodden-Turner (CDTT) \cite{Kausar, Sharif2}, and firstly considered in \cite{gCDTT} (see also \cite{Nojiri:2017opc}, \cite{Nojiri:2021iko} for applications of generalised Nojiri-Odintsov HDE to modified gravity):
\begin{equation}
\label{eq:f(R)}
f(R)= R+\lambda R^2- \sigma \frac{\mu}{R},
\end{equation}
where $\lambda$ and $\mu$ are small positive constants and $\sigma \pm 1$. At $\mu=0$ we obtain the usual Starobinsky model. \textbf{Below we will refer to this gravity model as the Nojiri-Odintsov gravity model.}

For convenience, let us pass to the variable $z$ instead of the time variable $z = {a(0)}/{a}-1$. For past times variable $z$ has the sense of a redshift. Then the time derivative ${d}/{dt}$ is related to ${d}/{dz}$ by the following relation: 
\begin{equation}
\label{eq:05}
\dfrac{d}{dt} = -H(z+1)\frac{d}{dz}.
\end{equation}
The first and second derivatives of the function $f(R)$ on argument are 
\begin{equation}
\label{eq:f(R)2}
{f}'(R) = 1+2\lambda R+ \sigma \frac{\mu}{R^2}, \quad
{f}''(R) = 2\lambda -\sigma \frac{2\mu}{R^3},
\end{equation}

By substituting the expressions (\ref{eq:f(R)}), (\ref{eq:f(R)2}) into equation (\ref{friedmann00}) and passing to variable $z$, we obtain the equation:

\begin{equation}
\frac{d^2H}{dz^2} = - \frac{f'(R)}{6H(z+1)^2f''(R)} + \frac{1}{18H^3(z+1)^2f''(R)}\left ( \rho + \frac{Rf'(R)-f(R)}{2} \right ) + \nonumber 
\label{eq:13}
\end{equation}
$$
+ \frac{1}{H(z+1)}\left ( \left ( \frac{dH}{dz} \right )^2 (z+1)- 3H\frac{dH}{dz} \right ).
$$
Supplementing (\ref{eq:13}) with the formula for the total energy density (\ref{eq:10}) and the equation for the variable $\tilde{L}$: 
$$\frac{d\tilde{L}}{dz}=\frac{1}{H}$$
we obtain a system of equations whose solution for given initial conditions determines the cosmological evolution of the Universe. 

\section{Analysis of the model}
 
For integration of equations describing the cosmological evolution in modified gravity, it is necessary to specify not only the initial scale factor and Hubble parameter $H(0)$ (Einstein-Friedmann equations have the second order on derivatives $a$), but also the second derivative of the scale factor $a$ on time at $t=0$. This is equivalent to the initial condition, imposed on $\dot{H}$ at $t=0$. In the analysis below (for $\lambda\neq 0$), we assume that $\ddot{a}(0)$ has the same value as in the $\Lambda$CDM model with $\Omega_{\Lambda} = 0.72$. This value is determined from the equation for $\dot{H}$ in Friedmann cosmology:
$$
\dot{H}=-\frac{1}{2}\left(\rho+p\right).
$$
As an initial density condition, we can take the values $\rho(0)=\rho_d(0) + \rho_m(0) = 3H_{0}^{2}$. Given that $p_{\Lambda}=-\rho_{\Lambda}$ and $p_{m}=0$, we obtain for $\Omega=0.72$ that
$\dot{H}(0)=-0.42 H_{0}^{2}$.
In terms of the $z$ variable, this condition means that $dH/dz=0.42 H_{0}^{2}$ for z=0. For $\lambda = \mu = 0$ $\dot{H}(0)$ is determined by parameter $C$ and $\Omega$. If $\gamma=1$ current equation-of-state parameter $w=p_{d}/\rho_{d}$ is
$$
w=-\frac{1}{3}-\frac{2}{3}\frac{\sqrt{\Omega}}{C}
$$
and therefore 
$$
\dot{H}(0)=-\frac{1}{2}\left(1-\Omega + \frac{2}{3}\left(1-\frac{\sqrt{\Omega}}{C}\right)\Omega\right)
$$
and slightly differs from value $0.42 H_{0}^{2}$ in $\Lambda$CDM cosmology. Let us analyze the case when in the expression (\ref{eq:f(R)}) $\mu = 0$. For $\lambda \neq 0$ in the future the Hubble parameter $H(t)$ begins to oscillate around that dependence which is observed for the holographic dark energy model in GR. The amplitude of these oscillations increases with time and value of $\lambda$. The analysis of the time derivative of the Hubble parameter $\dot{H}$ is also interesting. In GR the value of $H$ tends to a constant value, which corresponds to the fact that $\dot{H}\rightarrow 0$, and the derivative tends to zero from below. And for $\lambda \neq 0$ the function $\frac{dH}{dt}$ oscillates around zero at long times. The same future oscillations are observed for the scalar curvature $R$. In the past, the holographic dark energy model at $\lambda \neq 0$ behaves also differently in comparison with GR: the Hubble parameter and $|\dot{H}|$ increase not so strongly with time. It is interesting to note that relation between $H^2$ and $\dot{H}$ changes so that scalar curvature oscilates near the zero in past in modified gravity. In a case of Friedmann cosmology $R$ strongly increases and tends to $\infty$ which corresponds to big bang singularity. 

\begin{figure}
\centering 
    \includegraphics[scale=0.36]{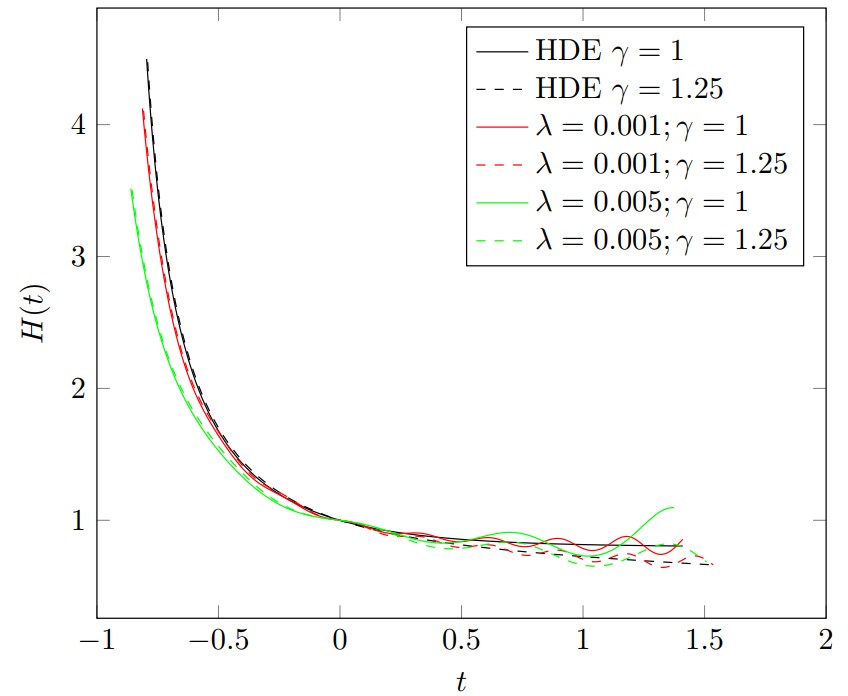}\includegraphics[scale=0.29]{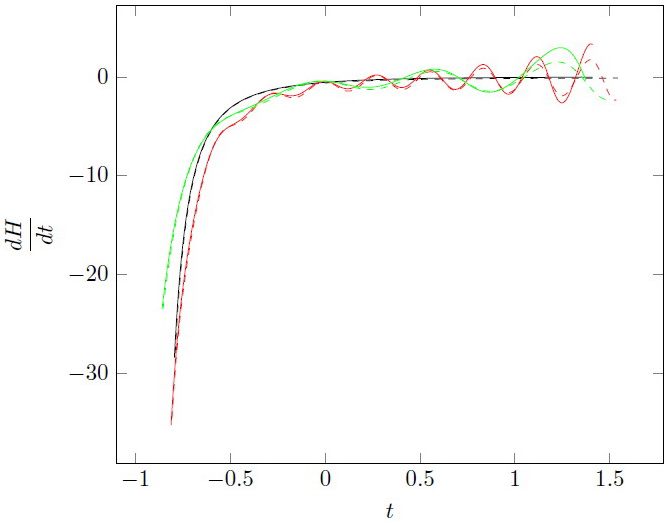}\\
    \includegraphics[scale=0.36]{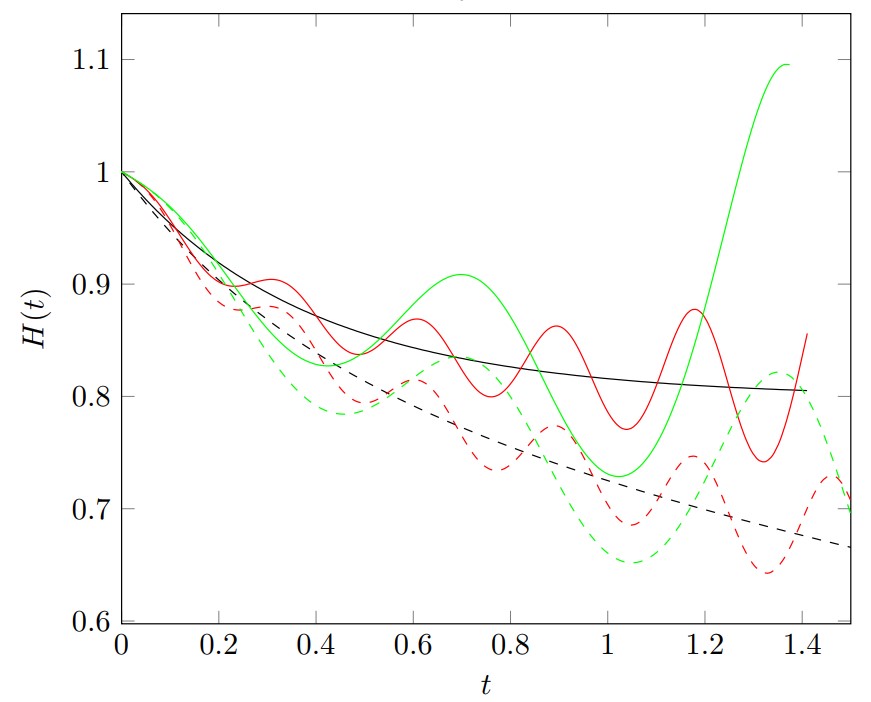}\includegraphics[scale=0.29]{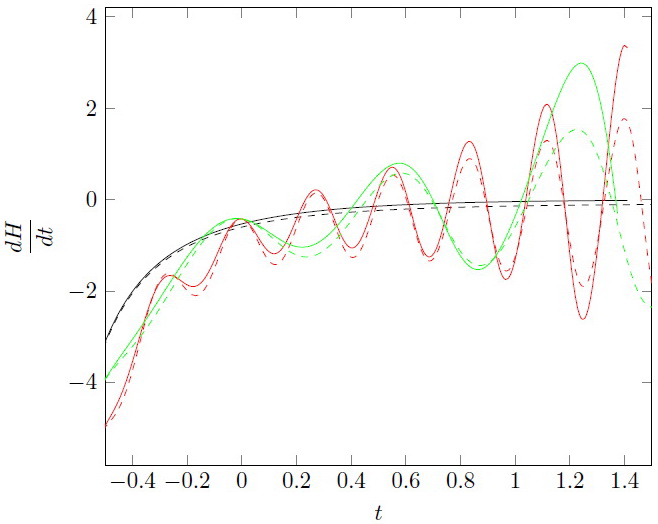}\\
    \includegraphics[scale=0.29]{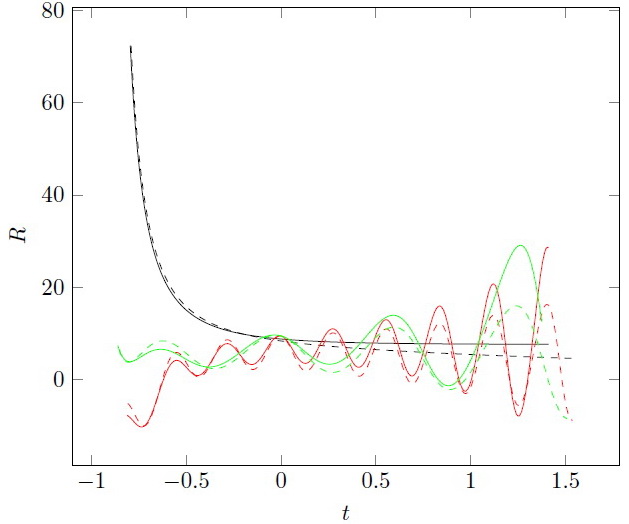}\includegraphics[scale=0.36]{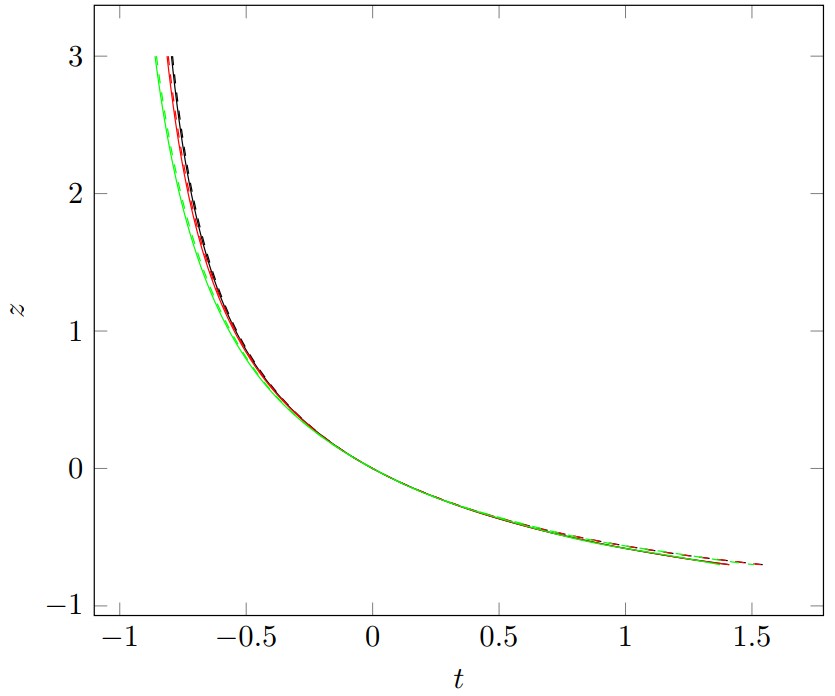}\\
    \caption{Dependence of the Hubble parameter $H$ and the derivative of the Hubble parameter $\dot{H}$ with time in the past and future (top panel) for $\mu = 0$, $C = 1$ and $\Omega = 0.72$. The Hubble parameter is measured in units of $H_{0}\equiv H(0)$ and time in units of $1/H_{0}$. Derivative of $H$ therefore is given in units of $H_{0}^2$. On the middle panel $H$ and $\dot{H}$ are given in more detail for a particular time interval. The bottom panel shows the dependence of the scalar curvature $R$ from time (left) and the relation between the parameter $z=1/a-1$ and time (right).}
    \label{fig1}
\end{figure}

\begin{figure}[ht]
\centering 
    \includegraphics[scale=0.36]{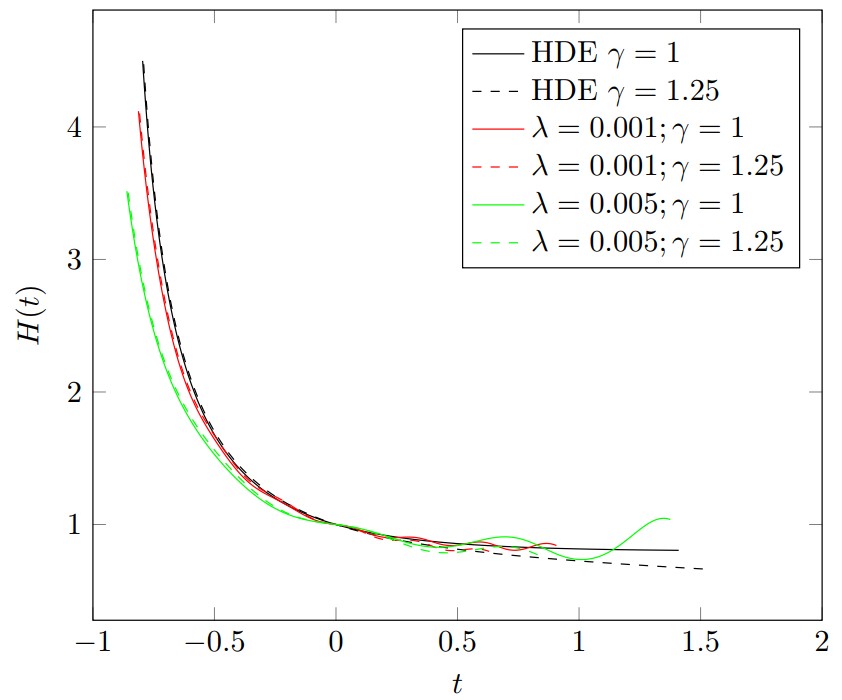}\includegraphics[scale=0.29]{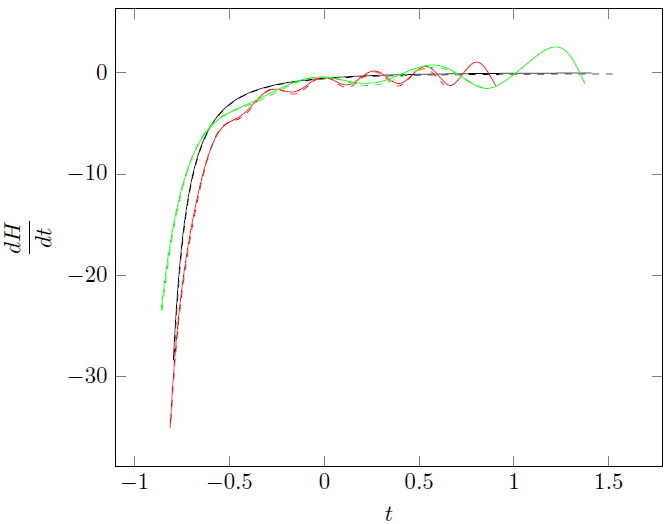}\\
    \includegraphics[scale=0.36]{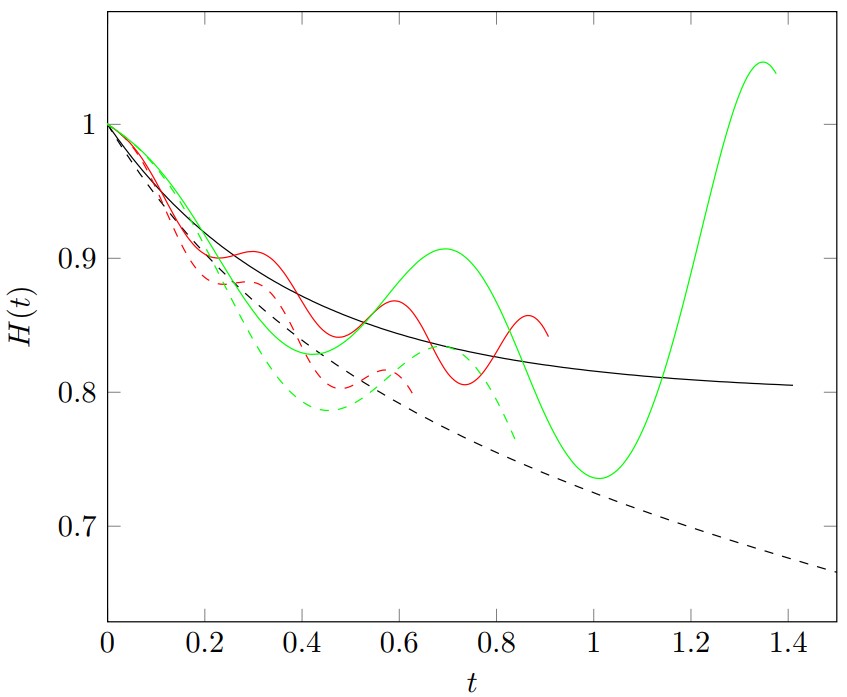}\includegraphics[scale=0.29]{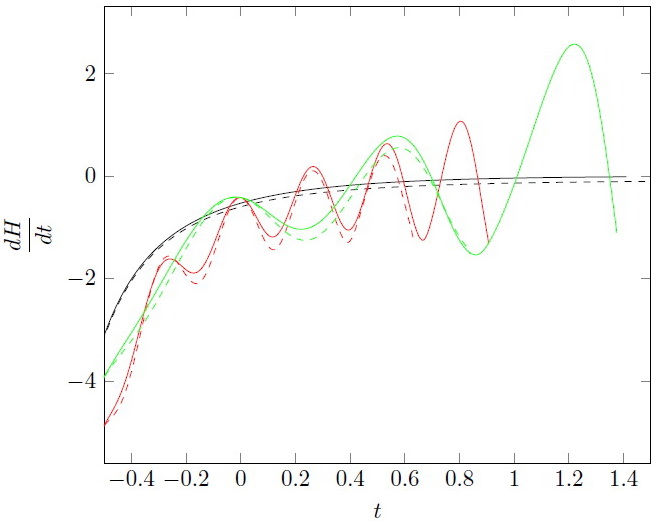}\\
    \includegraphics[scale=0.29]{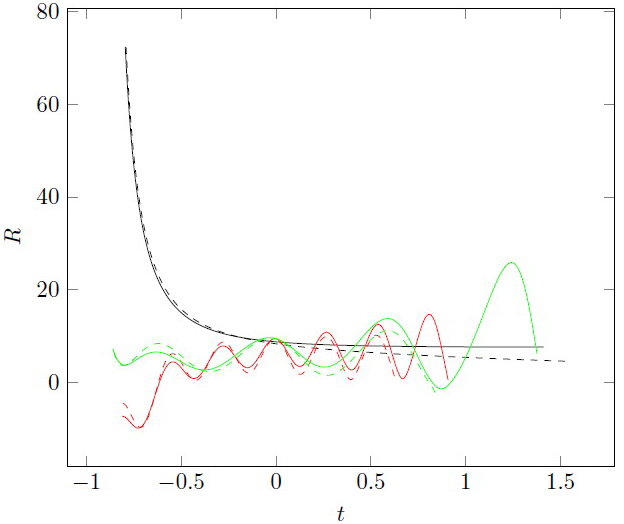}\includegraphics[scale=0.36]{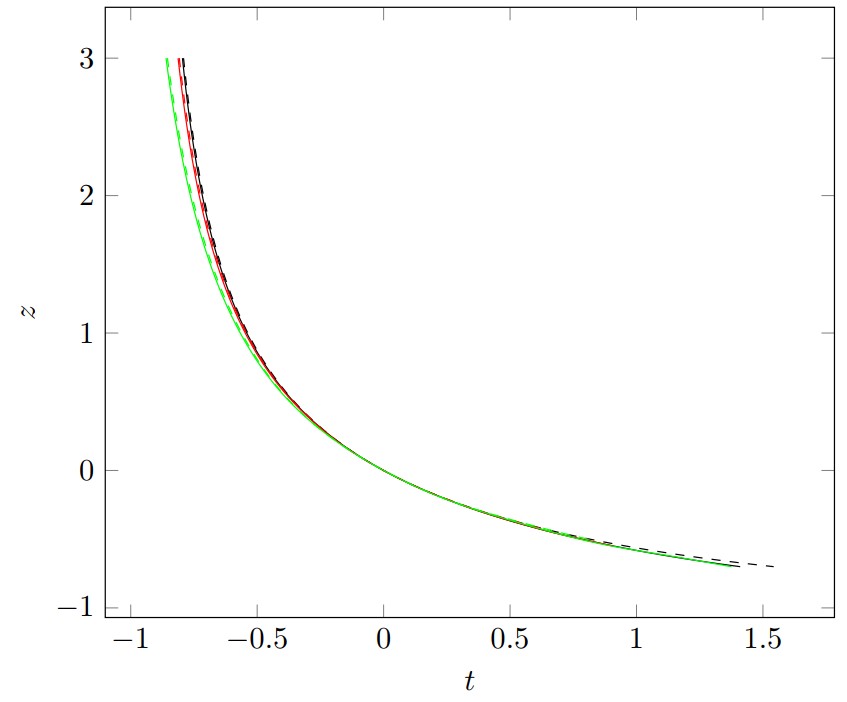}\\
    \caption{Same as the figure \ref{fig1}, but for $\mu = 0.001$, $C = 1$, $\Omega = 0.72$, $\sigma = 1$.}
    \label{fig2}
\end{figure}

\begin{figure}[ht]
\centering 
    \includegraphics[scale=0.36]{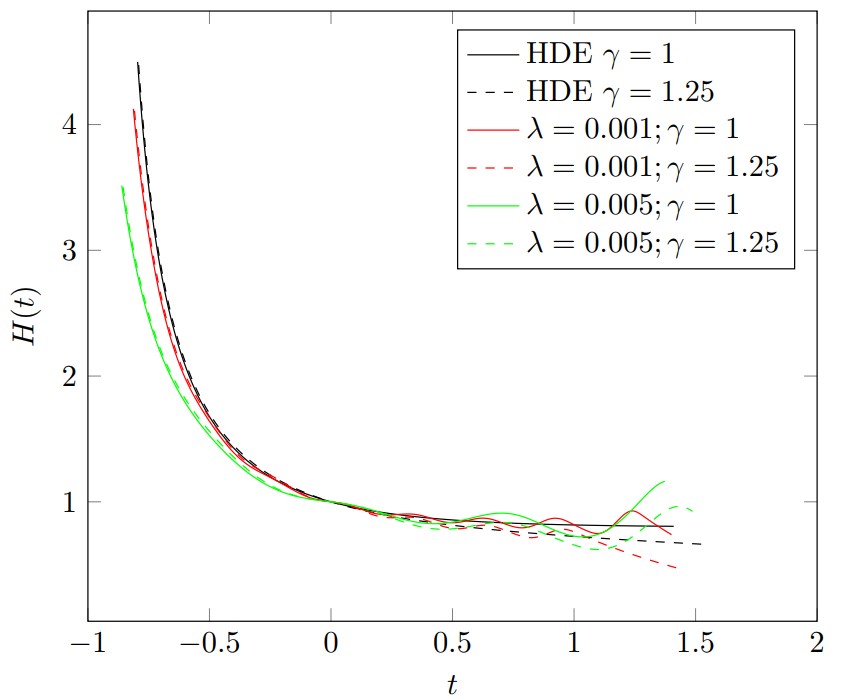}\includegraphics[scale=0.29]{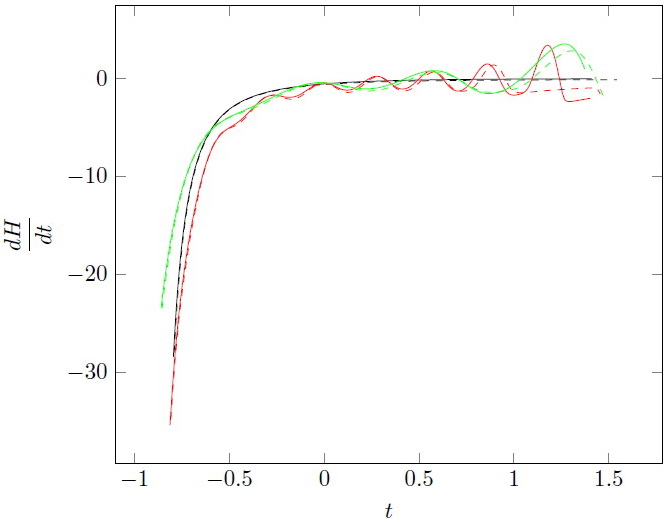}\\
    \includegraphics[scale=0.36]{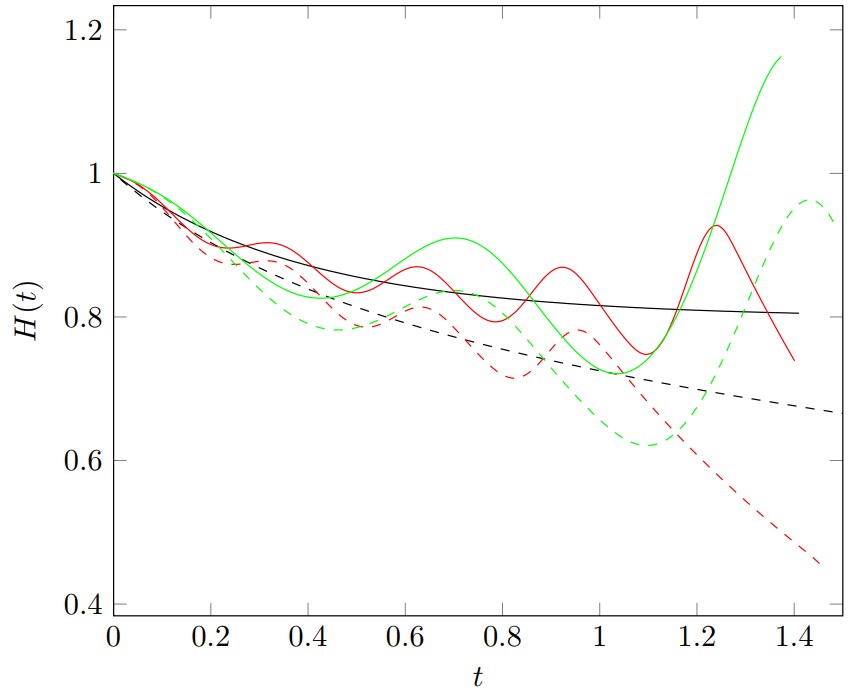}\includegraphics[scale=0.29]{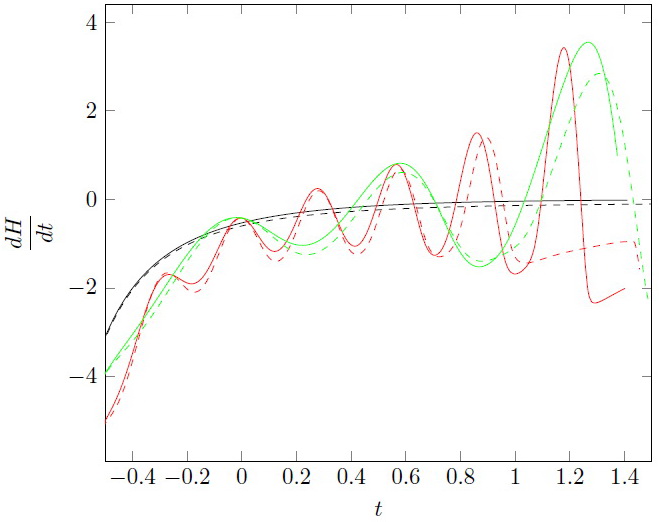}\\
    \includegraphics[scale=0.29]{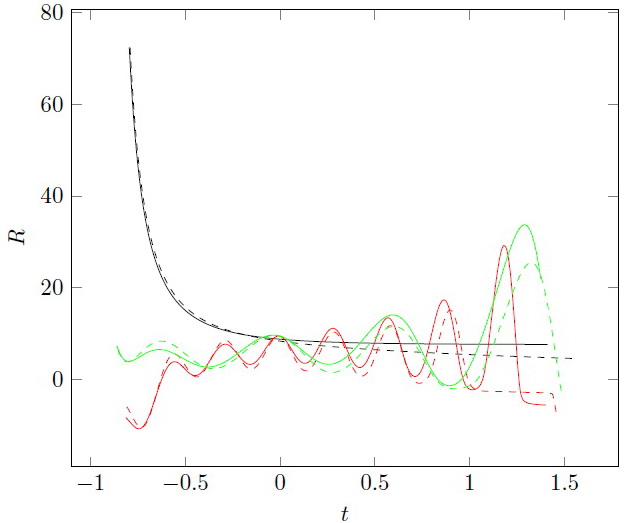}\includegraphics[scale=0.36]{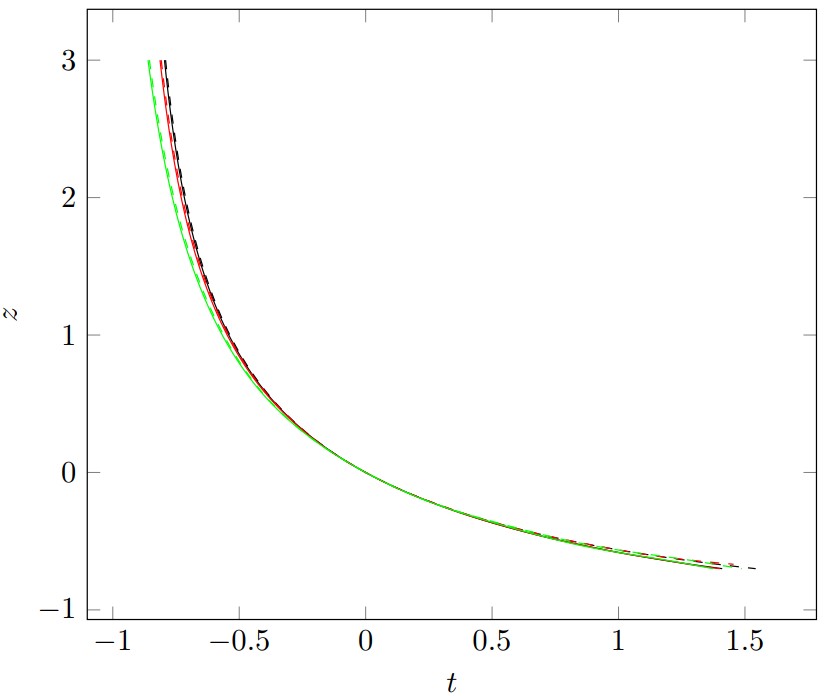}\\
    \caption{Same as the figure \ref{fig1}, but for $\mu = 0.001$, $C = 1$, $\Omega = 0.72$, $\sigma = -1$.}
    \label{fig3}
\end{figure}

\begin{figure}[ht]
\centering 
    \includegraphics[scale=0.36]{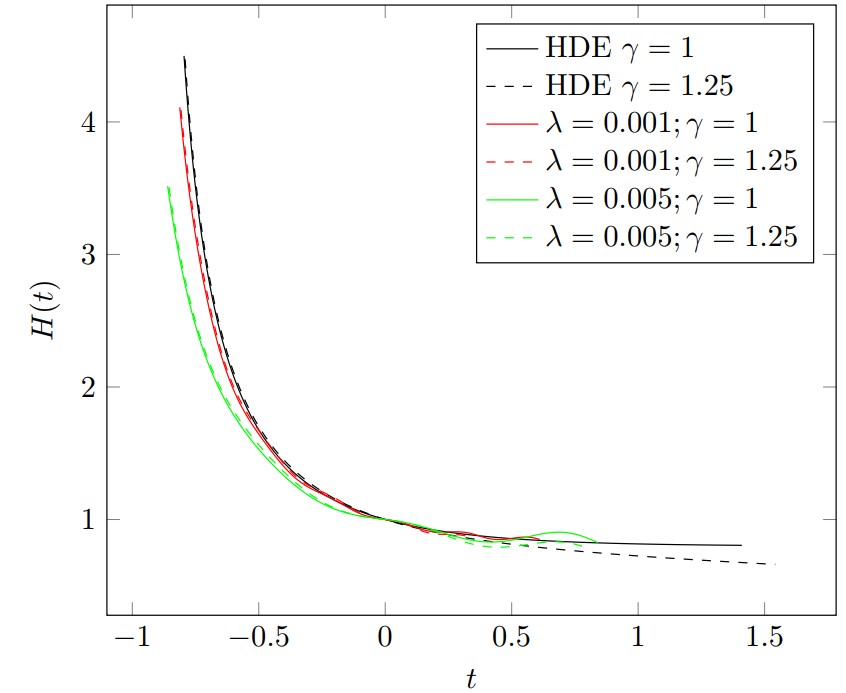}\includegraphics[scale=0.29]{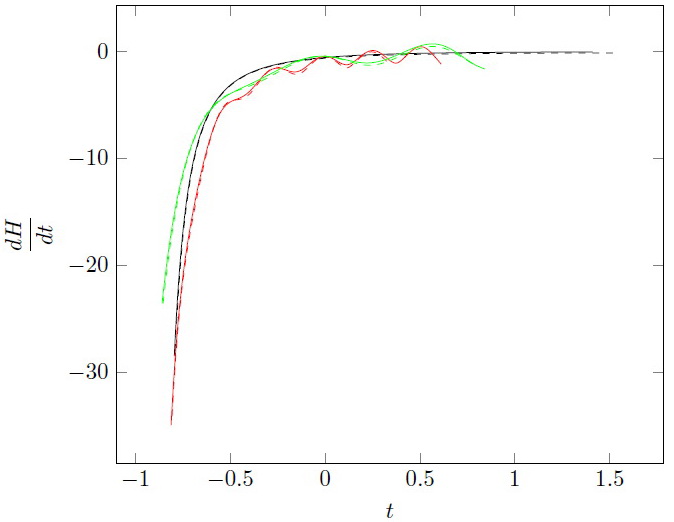}\\
    \includegraphics[scale=0.36]{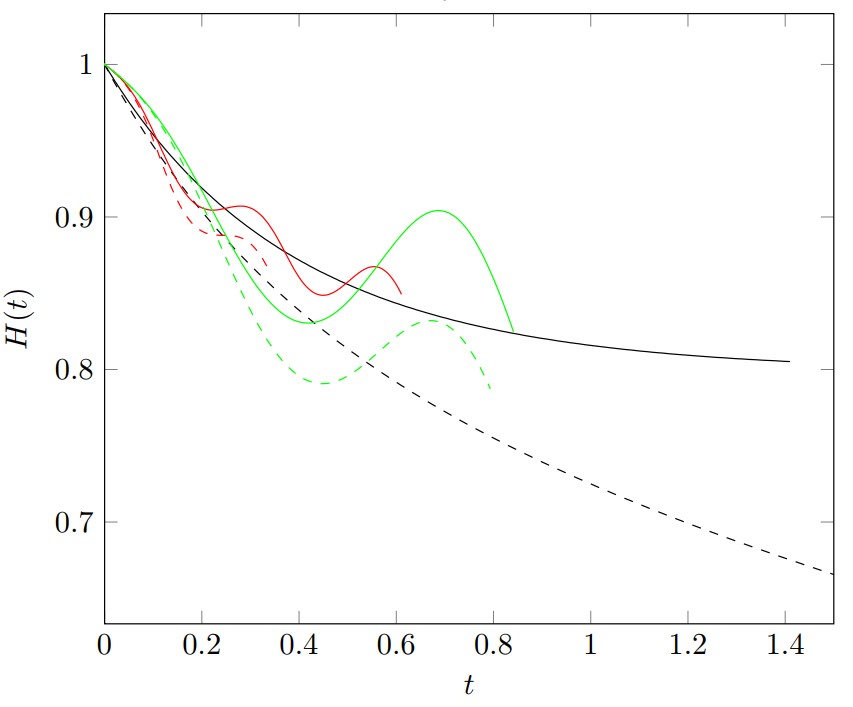}\includegraphics[scale=0.29]{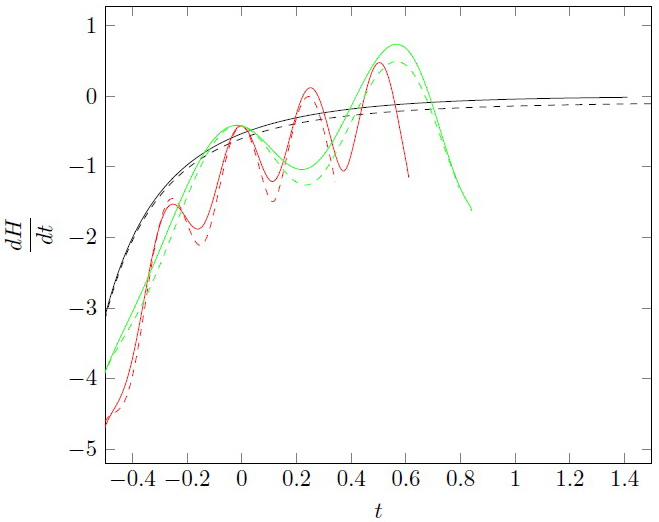}\\
    \includegraphics[scale=0.29]{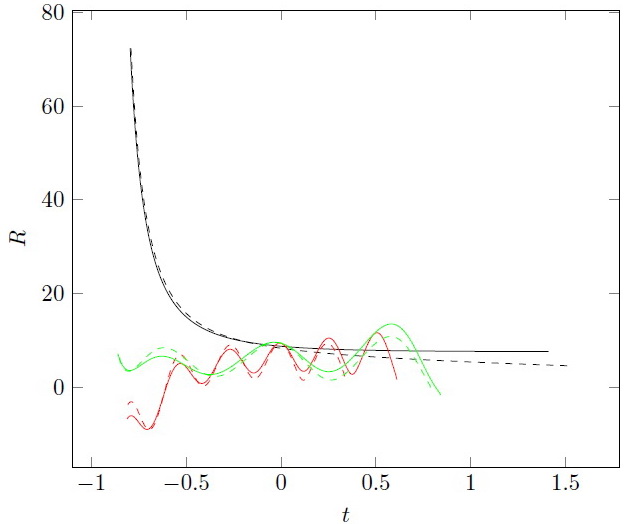}\includegraphics[scale=0.36]{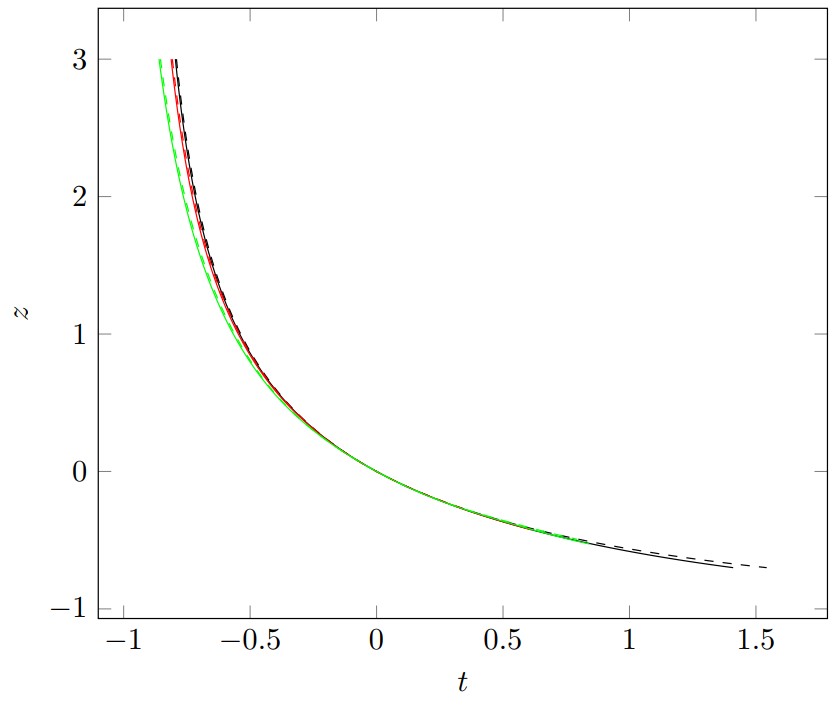}\\
    \caption{Same as in figure \ref{fig1}, but for $\mu = 0.003$, $C = 1$, $\Omega = 0.72$, $\sigma = 1$.}
    \label{fig4}
\end{figure}

\begin{figure}[ht]
\centering 
    \includegraphics[scale=0.36]{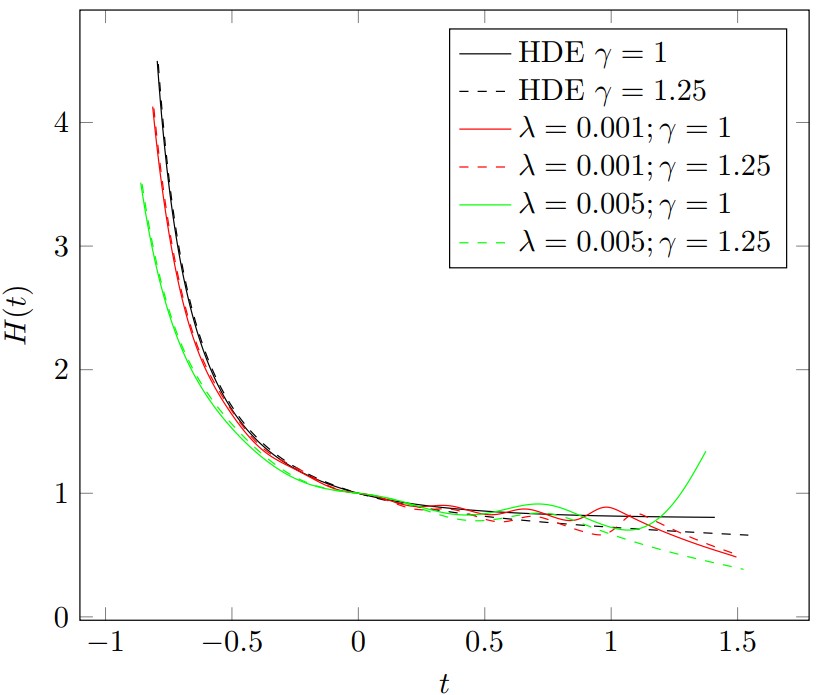}\includegraphics[scale=0.29]{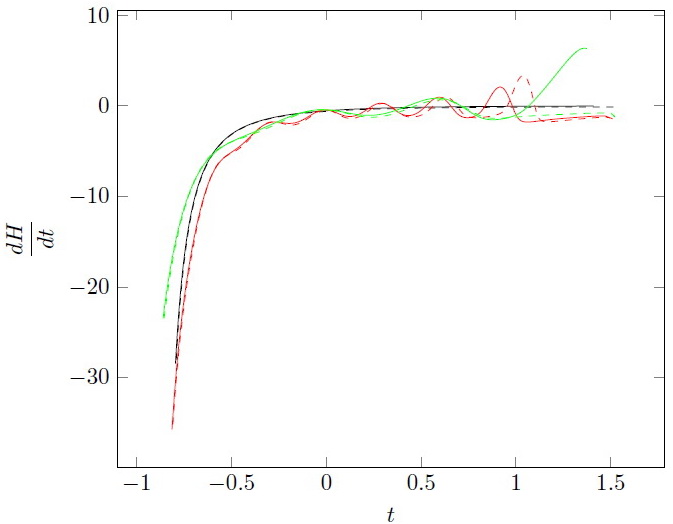}\\
    \includegraphics[scale=0.36]{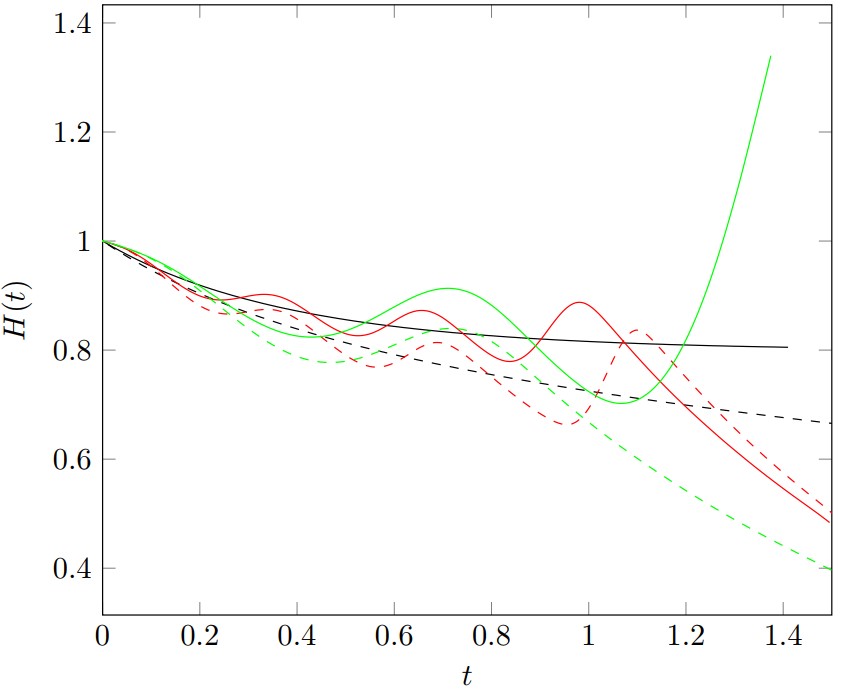}\includegraphics[scale=0.29]{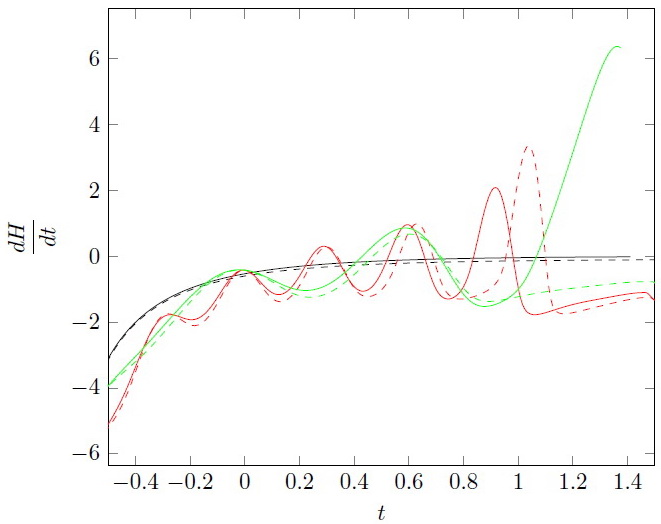}\\
    \includegraphics[scale=0.29]{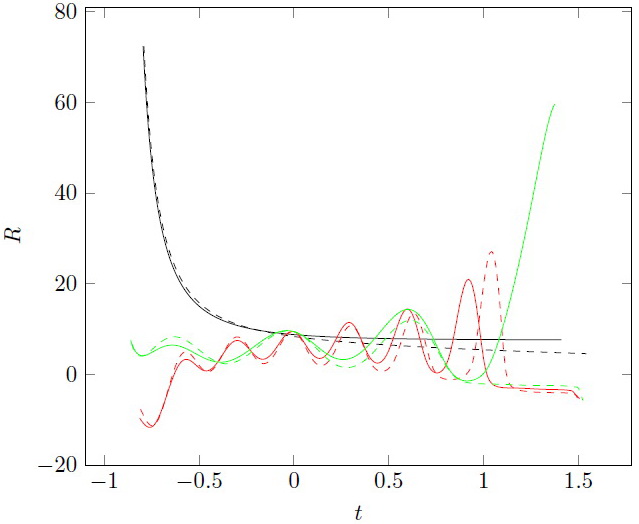}\includegraphics[scale=0.36]{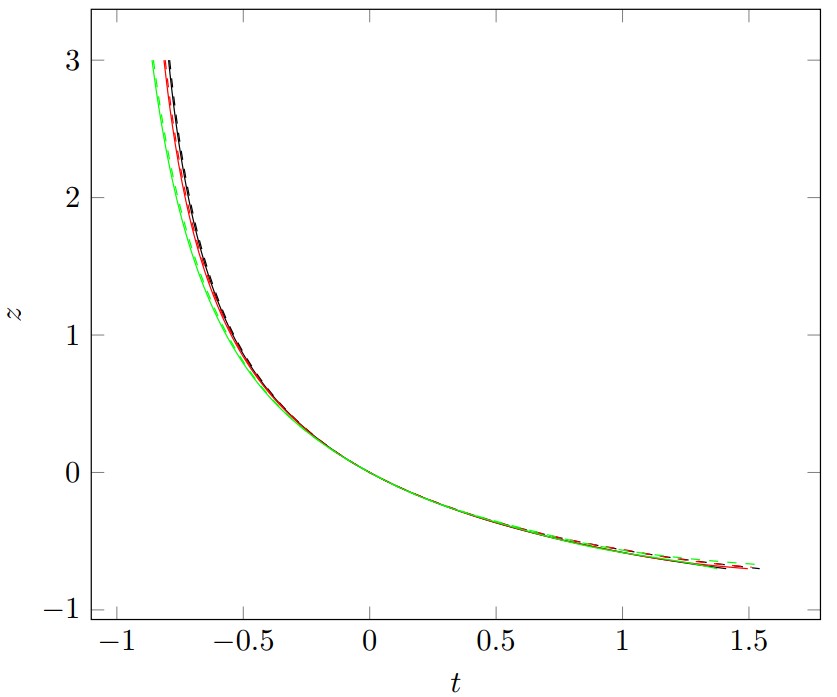}\\
    \caption{Same as in figure \ref{fig1}, but for $\mu = 0.003$, $C = 1$, $\Omega = 0.72$, $\sigma = -1$.}
    \label{fig5}
\end{figure}

If $\mu > 0$, the resulting dependences are very similar to the case when $\mu=0$. For $\mu=0.001$ the amplitude of oscillations of $H$ becomes smaller compared to the case of pure $R^2$-gravity (see Fig. 2). The main peculiarity is that singularities appear in the future, corresponding to the zero-point of the second derivative $f''(R)$. For $\sigma=-1$ the amplitude of the Hubble parameter oscillations, on the contrary, increases (Fig. 3). When the value of $\mu$ increases, the amplitude of oscillations of the Hubble parameter decreases for $\sigma=+1$ and increases for $\sigma=-1$. 

\begin{table}[ht]
\begin{center}
\begin{tabular}{|c||c|c|c|c|c|}
\hline
\backslashbox{$\lambda$}{$\sigma\mu $} & -0.003 & -0.001 &  0 & 0.001 & 0.003 \\
\hline
\multicolumn{6}{|c|}{$\gamma = 1$} \\
\hline
0.001 & 1.5005 & 1.4696 & 2.3678 & 0.9069  & 0.6107 \\
\hline
0.005 & 2.0362 &  1.6923 & 1.9583 & 1.4120  & 0.8412  \\
\hline
\multicolumn{6}{|c|}{$\gamma = 1.25$} \\
\hline
0.001 & 1.5112 & 1.4551 & 3.2991 &  0.6302 & 0.3409 \\
\hline
0.005 & 1.5234 & 1.7937 & 2.6955 & 0.8382 & 0.7934 \\
\hline
\end{tabular}
\caption{\label{tab:singularity}Time to the final singularity for different $\mu$ and $\lambda$ ($C = 1$, $\Omega = 0.72$).}
\end{center}
\end{table} 

We calculated the time before the singularity for these models (see Table I).

\begin{figure}[ht]
\centering 
    \includegraphics[scale=0.29]{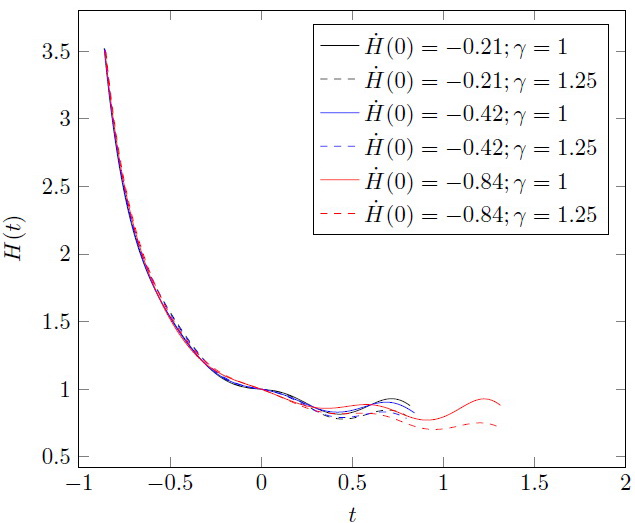}\includegraphics[scale=0.29]{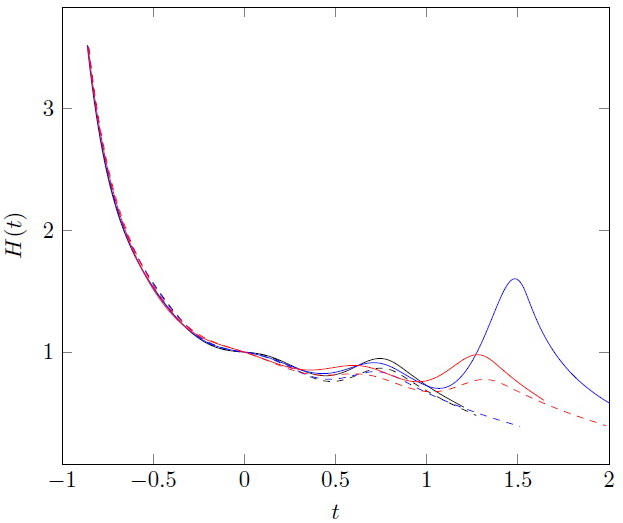}\\
    \includegraphics[scale=0.29]{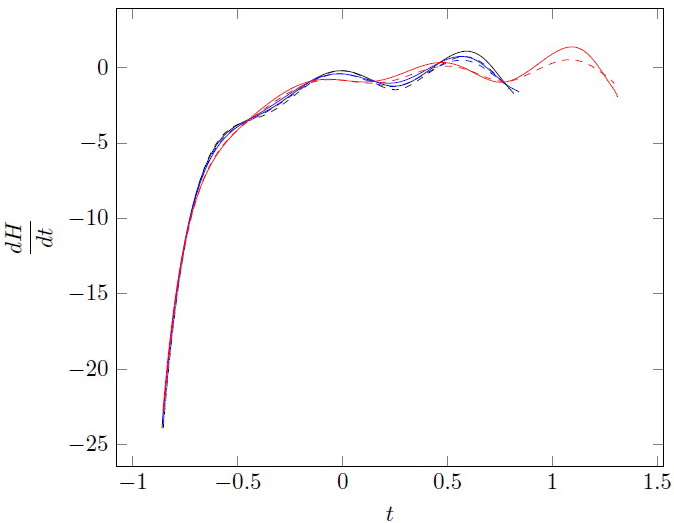}\includegraphics[scale=0.29]{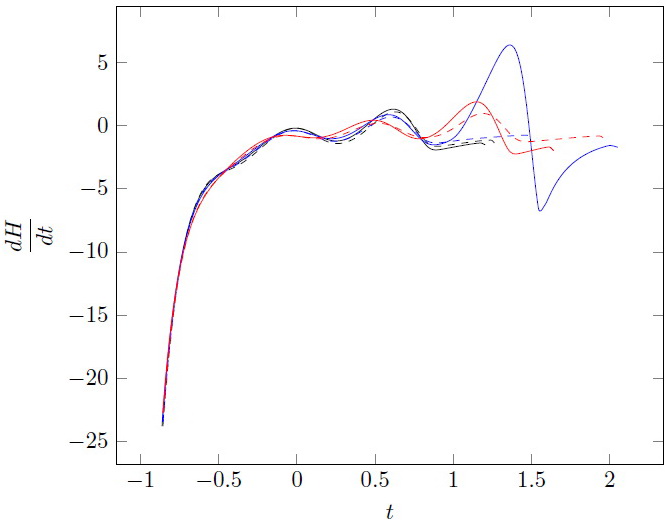}\\
    \caption{Dependence of $H$ and $\dot{H}$ for various initial condition on $\dot{H}(0)$ (in units of $H_{0}^{2})$ for $C = 1$, $\Omega = 0.72$. On left panel $\mu = 0.003$, $\lambda = 0.005$, $\sigma=1$, on right panel $\mu =  0.003$, $\lambda=0.005$, $\sigma=-1$.}
    \label{fig6}
\end{figure}

Let's consider the cosmological evolution for different values of $\dot{H}(0)$. For the $\Lambda$CDM model at $\Omega_{\Lambda} = 0.72$ the value of $\dot{H}_{\Lambda}(0)= - 0.42 \cdot H^{2}_{0}$. In units of $H_{0}^{2}$ therefore $\dot{H}(0) = - 0.42$. As mentioned above, in the $f(R)$-gravity model $\dot{H}(0)$ should be given as an initial condition for solving the equation. We consider several values for $\dot{H}_{0}$ = $-\dot{H}_{\Lambda (0)}; -2\cdot\dot{H}_{\Lambda}(0); -0.5\cdot\dot{H}_{\Lambda} (0)$. The calculations show that the evolution in the past weakly depends on the initial condition imposed on $\dot{H}(0)$ (see fig. \ref{fig6}), but the time before future singularity is significantly determined by this condition (Table \ref{tab:singularity1}).

Can these models be consistent with observational data? We analyzed this question by comparing the canonical $\Lambda$CDM model and the holographic dark energy model with $\lambda\neq 0$ and $\mu \neq 0$. Usually the following observational data tests are frequently used for cosmological models:

(i) relationship between apparent magnitude and redshift for Ia supernova from the Supernova Cosmology Project,

(ii) dependence of the Hubble parameter from redshift obtained from space chronometry and baryonic acoustic oscillation data.

Let's consider these observational limits in detail. Apparent magnitude $\mu(z)$ of star for a redshift $z=a_{0}/a-1$ is  
\be 
\mu(z)=\mu_{0}+5\log D(z)\, , 
\ee 
where $D_{L}(z)$ is the photometric distance defined by relation (for spatially flat universe)
\be
\label{DLSC} D_{L}(z)=\frac{c}{H_{0}}(1+z)\int_{0}^{z} h^{-1}(z)d
z, \quad h^{2}(z)=\rho(z)/\rho_{0}. \ee 
Here $c$ is the speed of light. The best fit for SNe Ia is given in the framework of $\Lambda$CDM cosmological model. For this model $h(z)$ is
\be 
h(z)=(\Omega_{m}(1+z)^{3}+\Omega_{\Lambda})^{1/2}.
\ee 
Here $\Omega_{m}$ is the matter density fraction, and $\Omega_{\Lambda}=1-\Omega_{m}$ is the vacuum energy fraction. The constant value $\mu_{0}$ depends from the current value of the Hubble parameter:
$$
\mu_{0}=42.384-5\log h,\quad h=H_{0}/100 \mbox{ km/s/Mpc}
$$
To analyze the SNe data, it is necessary to calculate the parameter $\chi^{2}$, which is defined by the standard relation
\begin{equation}
\chi^{2}_{SN}=\sum_{i}\frac{(\mu_{obs}(z_{i})-\mu(z_{i}))^{2}}{\sigma^{2}_{i}},
\end{equation}
where $\sigma_{i}$ is the corresponding $1\sigma$ error. We use known data for 580 SNe Ia from \cite{Amman}. The parameter $\mu_{0}$ is data-independent, so we can minimize $\chi^2 $ relative to $ \mu_{0} $. It should be noted that
\begin{equation}\label{chi}
\chi^{2}_{SN}=A-2\mu_{0}B+\mu_{0}^{2}C,
\end{equation}
where
$$
A=\sum_{i}\frac{(\mu_{obs}(z_{i})-\mu(z_{i};\mu_{0}=0))^{2}}{\sigma^{2}_{i}},
$$
$$
B=\sum_{i}\frac{(\mu_{obs}(z_{i})-\mu(z_{i}))}{\sigma^{2}_{i}},\quad C=\sum_{i}\frac{1}{\sigma^{2}_{i}}.
$$
The value of $\chi$-square (\ref{chi}) has a minimum at $\mu_{0}=B/C$, and this minimum is
$$
\bar{\chi}_{SN}^{2}=A-B^{2}/C.
$$
One can minimize $\bar{\chi}_{SN}^{2}$ instead of ${\chi}_{SN}^{2}$, and calculate corresponding optimal value of $H_{0}$. For 580 SNe samples and $\Lambda$CDM model minimal value $\bar{\chi}_{SN}^{2}=553.231$ for $\Omega=0.722$ and $H_{0}=70.05$ km/s/Mpc. 

For measuring the Hubble parameter for different redshifts $ z $ there are different methods. The lot of of data are obtained using the cosmic chronometry. The Hubble parameter depends from the differential age of the Universe as a function of redshift:
$$
dt=-\frac{1}{H}\frac{dz}{1+z}.
$$
Measurements of $ dz/dt $ (and, as a consequence, measurements of $H(z)$) are possible due to that absolute age data for galaxies is determined by fitting stellar population models. Results of these measurements are given in \cite{Zhang}, \cite{Simon}, \cite{Moresco2}, \cite{Moresco3}, \cite{Stern}, \cite{Ratsimbazafy}. There are also three correlated $H(z)$ measurements from the radial BAO signal \cite{Alam} in the galaxy distribution and two values for the large redshift ($z=2.34$ and $2.36$) measured from the BAO signal in the Lyman-alpha forest distribution \cite{Delubac}, \cite{Font-Ribera}. These 36 measurements of Hubble parameter $H(z)$ are listed in Table II.

\begin{table}
\label{Table1}
\begin{centering}
\begin{tabular}{|c|c|c||c|c|c|c|}
  \hline
  $z$ & $H_{obs}(z)$ & $\sigma $ & $z$ & $H_{obs}(z)$ & $\sigma$  \\
      & km s$^{-1}$ Mpc$^{-1}$        &  km s$^{-1}$ Mpc$^{-1}$ & & km s$^{-1}$ Mpc$^{-1}$        &  km s$^{-1}$ Mpc$^{-1}$     \\
  \hline
  0.070 & 69 & 19.6 & 0.480 & 97 & 62 \\
  0.090 & 69 & 12 & 0.510 & 90.8 & 1.9\\
  0.120 & 68.6 & 26.2 &  0.593 & 104 & 13\\
  0.170 & 83 & 8  & 0.610 & 97.8 & 2.1\\
  0.179 & 75 & 4 & 0.68 & 92 & 8\\
  0.199 & 75 & 5 & 0.781 & 105 & 12\\
  0.200 & 72.9 & 29.6 &  0.875 & 125 & 17\\
  0.270 & 77 & 14 & 0.880 & 90 & 40 \\
  0.280 & 88.8 & 36.6 & 0.900 & 117 & 23 \\
  0.352 & 83 & 14 & 1.037 & 154 & 20 \\
  0.38 & 81.9 & 1.9 & 1.300 & 168 & 17 \\
  0.3802 & 83 & 13.5 & 1.363 & 160 & 33.6\\
  0.400 & 95 & 17 & 1.430 & 177 & 18 \\
  0.4004 & 77 & 10.2 & 1.530 & 140 & 14 \\
  0.4247 & 87.1 & 11.2 & 1.750 & 202 & 40 \\
  0.4497 & 92.8 & 12.9 & 1.965 & 186.5 & 50.4\\
  0.470 & 89 & 50 & 2.34 & 223 & 7\\
  0.4783 & 80.9 & 9 & 2.36 & 227 & 8\\
  \hline
\end{tabular}
\caption{The dependence of the Hubble parameter $H(z)$ from observations used in our analysis of THDE model in gCDTT gravity}
\end{centering}
\end{table}

The parameter $\chi^{2}_{H}$ is
\begin{equation}
\chi^{2}_{H}=\sum_{i}\frac{(H_{obs}(z_{i})-H(z_{i}))^{2}}{\sigma^{2}_{i}}.
\end{equation}
We can also perform averaging over the unknown parameter $H_{0}$. We obtain that
$$
\chi^{2}_{H}=A_{1}-2B_{1}H_{0}+H_{0}^{2}C_{1},
$$
$$
A_{1}=\sum_{i}\frac{H_{obs}(z_{i})^{2}}{\sigma^{2}_{i}},\quad B_{1}=\sum_{i}\frac{h(z_{i})H_{obs}(z_{i})}{\sigma^{2}_{i}},\quad
$$
$$
C_{1}=\sum_{i}\frac{h(z_{i})^2}{\sigma^{2}_{i}}.
$$
For $ H_ {0} = B_{1}/C_{1} $ the parameter $\chi^{2}_{H} $ is minimal.
$$
\bar{\chi}_{H}^{2}=A_{1}-B_{1}^{2}/C_{1}.
$$
As in the case of supernova data, we can find a minimum of $\bar {\chi}_{H}^{2} $ instead of ${\chi}_{H}^{2}$. For $\Lambda$CDM model we have that $\bar{\chi}^{2}_{H}$ is minimal for $\Omega_{\Lambda}=0.737$ and equal to $19.262$. Corresponding $H_{0}$ is 70.32 km/s/Mpc. Therefore analysis of two data sets gives similar results for $H_{0}$ and optimal value of $\Omega_\Lambda$.

Analysis of THDE model in gCDTT gravity shows some interesting moments. Firstly observational data favor to $\gamma=1$ (canonical model of HDE). 
Secondly data set for Hubble parameter is described by holographic model better in comparison with $\Lambda$CDM model. For example if $\lambda=0.001$, $\mu=0.001$, $\sigma = 1$ we have following minima of $\bar{\chi}_{H}^{2}$ for various $\dot{H}(0)$:

17.097 ($H_0=67.89$ km/s/Mpc) for  $\Omega=0.717$, $\dot{H}(0)=-0.21 H_{0}^{2}$; 

16.955 ($H_0=67.64$ km/s/Mpc) for $\Omega=0.713$, $\dot{H}(0)=-0.42 H_{0}^{2}$;

16.799 ($H_0=67.13$ km/s/Mpc) for $\Omega=0.705$, $\dot{H}(0)=-0.84 H_{0}^{2}$.

But for these parameters SNe data are described worse: for given $\dot{H}(0)$ there is a  significant discrepancy between optimal value of $\Omega$ from two data sets. However for some $\Omega$ and $\dot{H}(0)$ we obtained that SNe and Hubble data sets in general are fitted with same accuracy as for $\Lambda$CDM model. In particular for $\dot{H}(0)=-0.84 H_{0}^2$ we have that minimum of $\bar{\chi}^{2}_{SN}+\bar{\chi}^{2}_{H}$ is 572.6723 for $\Omega=0.745$. This corresponds to $\Lambda$CDM model with $\Omega_{\Lambda}=0.732$. Therefore one conclude that THDE model in frames of gCDTT gravity can be considered as quite realistic model of cosmological acceleration. 

\begin{table}[ht]
\begin{center}
\begin{tabular}{|c||c|c|c|}
\hline
\backslashbox{$\sigma \mu$}{$\dot{H}(0)$} & -0.21 & -0.42 &  -0.84 \\
\hline
\multicolumn{4}{|c|}{$\gamma = 1$} \\
\hline
-0.003 & 1.2041 & 2.0521 & 1.6429 \\
\hline
0.003 & 0.8162 & 0.8412 & 1.3122 \\
\hline
\multicolumn{4}{|c|}{$\gamma = 1.25$} \\
\hline
-0.003 & 1.2718 & 1.5116  & 1.9824 \\
\hline
0.003 & 0.7931 & 0.7970 & 1.2959  \\
\hline
\end{tabular}
\caption{\label{tab:singularity1} Time before the final singularity (in units of $1/H_0)$ for different $\mu$ and $\dot{H}(0)$ ($C = 1$, $\Omega = 0.72$, $\lambda = 0.005$). $\dot{H}(0)$ is given in units of $H_{0}^{2}$.}
\end{center}
\end{table} 

\begin{figure}[ht]
\centering 
    \includegraphics[scale=0.6]{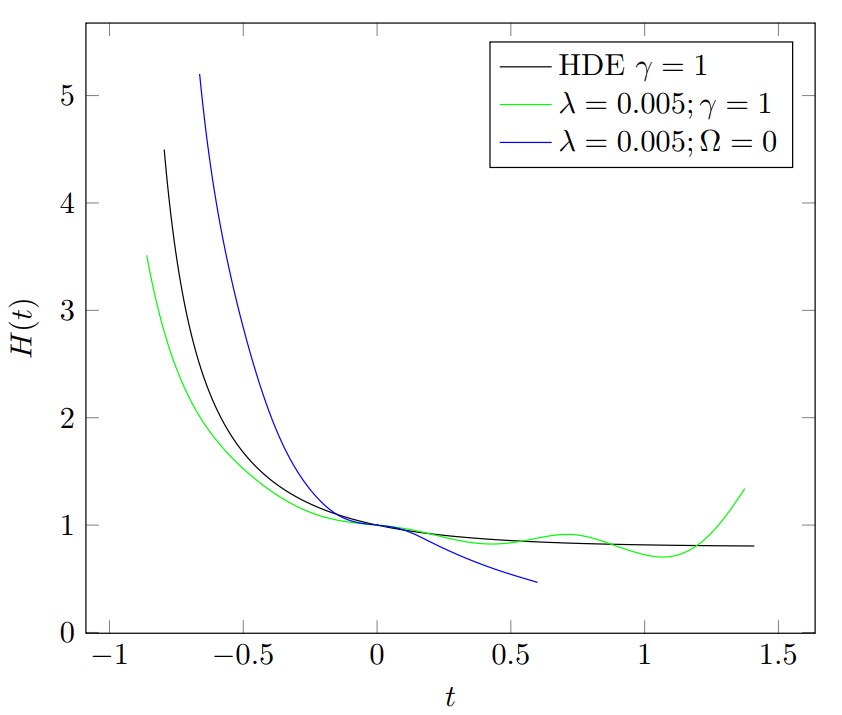}
    \caption{Dependence of the Hubble parameter for (i) the holographic dark energy model in Friedmann cosmology ($\Omega = 0. 72$), (ii) the model of holographic dark energy in Nojiri-Odintsov gravity ($\mu = 0.003$, $\sigma=-1$, $\lambda = 0.005$) with $\dot{H}(0)=-0.42$ (iii) and the model of Nojiri-Odintsov gravity with only matter for the same $\mu$, $\lambda$ and $\dot{H}(0)$.}
    \label{fig7}
\end{figure}

It is also interesting to compare the evolution of the Universe in the model of Nojiri-Odintsov gravity without dark energy and in the models with holographic dark energy against the background of GR and the considered model of gravity (see fig. 7). Calculations leads to conclusion that oscillations of Hubble parameter in future is specific feature of universe filled THDE in Nojiri-Odintsov gravity.  

\section{Concluding remarks}

We investigated the Tsallis holographic dark energy model with assumption of the Nojiri-Odintsov gravity model is valid. The equations describing the cosmological evolution in this case contain third derivative of the scale factor on time. Therefore this requires to impose initial condition on the second derivative $a$ (which is equivalent to the condition on $\dot{H}(0)$). The evolution of the universe is studied in detail for the case when $\dot{H}(0)$ coincides with the value in the standard cosmological model with $\Omega_\Lambda = 0.72$. Solutions have interesting feature namely Hubble parameter ``oscillates'' near dependence corresponding to THDE in General Relativity. The amplitude of this oscillations grows with time in future.  For $\mu\neq 0$ a future singularity arises corresponding to zero of second derivative of $f(R)$ for some $R$. The time before singularity, as determined by the value of $\dot{H}$ for the initial moment in time, can vary in wide limits. Dynamics of the universe in the past is not especially sensitive to this initial condition and is close to that in the model of holographic dark energy in the background of GTR (the differences appear only at times close to the initial singularity of the Big Bang). Our analysis shows that such models for some parameters can describe observational data for SN Ia and dependence $H(z)$ with sufficient accuracy especially for $\gamma=1$ and larger values of $\dot{H}$ in comparison with $\Lambda$CDM model. Also one note that $H(z)$ data are described better in frames of THDE on modified gravity backgroud. This means that the models considered by us can be quite realistic.

\section*{Acknowledgments}

This work was supported by Ministry of Education and Science
(Russia), project 075-02-2021-1748.

\end{document}